\documentclass{article}

\usepackage[dblblindworkshop,final]{neurips_2026}
\workshoptitle{AI-Native Academia}

\usepackage[utf8]{inputenc}
\usepackage[T1]{fontenc}
\usepackage{hyperref}
\usepackage{url}
\usepackage{booktabs}
\usepackage{microtype}
\usepackage{xcolor}
\usepackage{graphicx}
\usepackage{array}
\usepackage{tabularx}
\usepackage{multirow}
\usepackage{wrapfig}
\usepackage{enumitem}
\usepackage[all]{nowidow}
\setlist{nolistsep,leftmargin=*}

\newcommand{\Sref}[2][]{\hyperref[#2]{Sec.~\ref*{#2}#1}}
\newcommand{\Fref}[2][]{\hyperref[#2]{Fig.~\ref*{#2}#1}}
\newcommand{\Tref}[2][]{\hyperref[#2]{Tab.~\ref*{#2}#1}}
\newcommand{\Eref}[2][]{\hyperref[#2]{Eq.~\ref*{#2}#1}}
\newcommand{\Aref}[2][]{\hyperref[#2]{Appx.~\ref*{#2}#1}}
\newcommand{\Dref}[2][]{\hyperref[#2]{Def.~\ref*{#2}#1}}
\hypersetup{%
    colorlinks,
    linkcolor={red!50!black},
    citecolor={green!80!black}
}

\title{Governing AI Research Through Peer Review:\\A Mixed-Methods Study of the Longitudinal\\Effects of Ethics Flags Across Resubmissions}
\author{%
Kento Nishi$^{1,2}$ \quad
Alec Laprevotte$^{2}$ \quad
Isaiah Bullock$^{2}$ \quad
Mfoniso M. Andrew$^{2}$
\\
{\small $^{1}$Massachusetts Institute of Technology \quad $^{2}$Harvard University}\\
{\small Correspondence: \texttt{knishi@mit.edu}}
}

\begin{document}
\maketitle

\begin{abstract}
Selective AI conferences have recently begun enforcing ethics review processes, with the goal of steering research towards safer and more responsible practices before publication. But do ethics flags actually steer research as intended? In this paper, we show that authors often revise the framing of their project following ethics flags rather than redirecting their underlying research agendas. We first study the longitudinal effects of ethics flags, tracing flagged ICLR~(International Conference on Learning Representations) submissions that were rejected or withdrawn into subsequent public resubmissions and tracking manuscript changes made after the review process ends when the original reviewers no longer oversee the project. We qualitatively classify these resubmissions into five categories based on what changed after review and find that in 83\% of 446 cases, authors leave the flagged concern unaddressed or revise the paper without changing the implicated methods or procedures. Then, we ask: if authors rarely change the research in response to ethics flags, what do they change instead? To answer this, we manually read reviews and rebuttals from 25 cases and interview authors about their rebuttal processes and resubmission decisions. We find that authors often make concessions during rebuttal while reviewers can still update their scores, but drop those concessions after rejection, when the original reviewers no longer have leverage. Interview participants describe publication changes as separate from research direction changes, calling review an ``editorial process'' that shapes ``what stories get seen'' and, in another case, saying peer reviews are ``mostly to filter out papers.'' Since authors more readily change what they publish than what they study or build, we recommend several policy changes, especially disclosure of prior ethics flags upon resubmission so accountability for unresolved concerns carries over.
\end{abstract}

\section{Introduction}
The Conference on Neural Information Processing Systems~(NeurIPS), the International Conference on Machine Learning~(ICML), and the International Conference on Learning Representations~(ICLR) have increasingly formalized research ethics through broader impact and ethics statements, submission checklists, and ethics flags~\citep{prunkl2021institutionalizing,neurips2026checklist,neurips2021retrospective,icml2023guidelines,iclr2023ethics}. Under these requirements, reviewers can object not only to technical claims, but also to how researchers collect data, evaluate harms, release artifacts, or anticipate misuse. These objections warrant attention, since choices about data, system design, and deployment distribute risks, encode values, and reshape social relations.~\citep{winner1980artifacts,nissenbaum1996accountability,selbst2019fairness}. Although researchers cannot be expected to foresee every downstream impact, \citet{stilgoe2013responsible} argue for anticipating and responding to plausible effects, while \citet{do2023important} distinguish unintended from unanticipated consequences and emphasize researchers' responsibility for foreseeable effects. Since major conferences shape which research receives attention and legitimacy, they share responsibility for what they accept and promote~\citep{hecht2021negative}. In practice, conferences exercise this responsibility through peer review, giving reviewers leverage to steer research towards safer and more responsible directions.

Yet reviewers gain this leverage very late in most research projects. By the time they evaluate a submission, authors have usually chosen the problem, collected the data, implemented the system, and completed most experiments. Since reviewers can update scores in response to rebuttals and author comments during the ``discussion period,'' authors are naturally incentivized to acknowledge and address the reviewers' concerns. But once a rejection decision is issued, the original reviewers no longer influence whether or how the work is disseminated moving forward. Unfortunately, reviewer judgments are notoriously noisy, so a new reviewer panel at the next conference may not flag the same unresolved concern. This creates room for performative concessions during the discussion period that are not necessarily carried into revisions for the next venue.

Ideally, authors who accept reviewer criticism would implement these recommendations before submitting their work again. If such steering occurs, we would expect clear changes across submissions: authors who acknowledge a privacy concern, for example, may remove identifiers, restrict data access, add an evaluation, or change a release decision. How often, then, do authors change their research in response to ethics flags after the original review ends? To answer this question, we follow rejected and withdrawn ICLR submissions with ethics flags into public resubmissions, compare each concern with the corresponding methods and procedures, and qualitatively classify each resubmission into one of five categories. We find that out of 446 resubmissions, only 17\% exhibit procedural or methodological changes addressing a previous concern. This raises another adjacent question: if authors rarely change the research, then what do they change instead? We cannot answer this from tracking papers alone, because comparing two versions lets us identify what authors changed but not why they changed it. That is, one may revise a method because they accepted criticism, or because the project evolved independently; it is also possible that they reject a criticism but still end up rewriting the paper for a different reason, like a change in target audience. To uncover the actual reasons for changes, or lack thereof, we read complete review histories for 25 cases and interview authors about their rebuttal and resubmission decisions. Our qualitative analysis of reviews and interviews together reveals that while authors frequently revise papers, they rarely cite reviewer-issued ethics flags as genuine reasons to redirect the underlying research.

\section{Related Work}
\textbf{Ethical reflection in technical research.} Sociotechnical systems and human-computer interaction researchers have long argued that technical design and evaluations can distribute social consequences~\citep{winner1980artifacts,nissenbaum1996accountability,selbst2019fairness,stilgoe2013responsible,karusala2024understanding}. \citet{hecht2021negative} propose requiring researchers to confront negative impacts during peer review; \citet{prunkl2021institutionalizing} argue broader impact requirements can institutionalize ethical reflection.

\textbf{Ethics review at top conferences.} In 2021, NeurIPS established new ``Ethics Guidelines'' and issued several conditional acceptances and one rejection on ethical grounds~\citep{neurips2021retrospective}. Following this, \citet{liu2022examining} examine discussions from that cycle. Later, ICLR 2023 program chairs reported screening 190 papers with ethics flags~\citep{iclr2023ethics}. Other work finds varying discussion of harms and limitations and difficulty translating ethical principles into practice~\citep{nanayakkara2021broader,ashurst2022ethics,pant2024ethics,morley2023operationalising}. These works examine isolated review cycles or ethical practice more broadly, but do not consider whether ethics reviews shape research and in what way.

\section{Longitudinal Analysis: Do Authors Change Research After Ethics Flags?}
To study the longitudinal effects of ethics flags, we need rejected or withdrawn projects that authors resubmit after the original review ends. Only ICLR publicly releases all submissions and reviews after decisions; NeurIPS and ICML release rejected papers only when authors opt in~\citep{iclr2026authorguide,neurips2026handbook,icml2026authorinstructions}. We therefore collect 47,893 ICLR submissions and 185,194 public reviews from OpenReview~(2021--2026). We search reviews for recorded instances of the ``Flag for ethics review'' checkbox, identifying 2,498 flagged submissions. Our 2021 snapshot lacks public flags, so counts begin in 2022 and rise from 107 to 1,100 in 2026 across six ethics categories. Of these, 1,857 were rejected or withdrawn, so we search them for resubmissions. We cannot simply match titles because authors often change a paper's title, abstract, coauthors, or scope. Accordingly, we query OpenAlex~\citep{priem2022openalex} with title variants, author surnames, and salient abstract terms, then rank candidates by title similarity, author and abstract overlap, and publication year. We~retrieve 6,618 candidates and retain 601 top matches. A retrieval score is insufficient to establish project identity, so we manually read each pair and retain 446 resubmissions.\footnote{Code for corpus construction and retrieval is available at \href{https://github.com/KentoNishi/neurips26-ai-ethics}{https://github.com/KentoNishi/neurips26-ai-ethics}.}

Even after identifying a resubmission, a version difference alone does not show steering, so we look for a necessary observable consequence: changes to the methods or procedures implicated by the ethics concern. For each pair, we locate the concern in the original review, compare the corresponding parts of the submission and resubmission, and qualitatively code each resubmission into one of five categories.\footnote{Manual coding was carried out by our team of four researchers, with regular meetings to calibrate labels.}  We record \emph{no observable change} when authors leave the concern unaddressed and \emph{mainly rhetorical change} when they change limitations, discussion of harms, or claims without changing the implicated technical work. We separately record changes to \textit{procedures}, \textit{methods}, or \textit{both}: procedures include consent, release, access, oversight, or governance; methods include data, models, evaluations, or experimental design. Rhetorical revisions can improve a paper, but they do not change the implicated methods or procedures and therefore do not count as research changes.

\setlength{\columnsep}{4pt}
\begin{wraptable}{r}{0.26\textwidth}
\vspace{-1em}
\centering
\scriptsize
\renewcommand{\arraystretch}{0.90}
\setlength{\tabcolsep}{2.0pt}
\begin{tabular}{@{}lr@{}}
\toprule
Change after the flag & Count \\
\midrule
No observable change & 217 \\
Mainly rhetorical & 153 \\
Procedures + methods & 57 \\
Primarily procedures & 14 \\
Primarily methods & 5 \\
\midrule
Total labeled & 446 \\
\bottomrule
\end{tabular}
\setlength{\abovecaptionskip}{2pt}
\setlength{\belowcaptionskip}{0pt}
\caption{Resubmissions.}
\label{tab:changes}
\vspace{-1.1em}
\end{wraptable}
In~\Tref{tab:changes}, we provide strong evidence that most authors leave flagged methods and procedures unchanged. In 217 of 446 resubmissions, authors make no observable change related to the concern; in another 153, they change mainly how they frame or discuss it. These two categories cover 370 resubmissions~(83.0\%). In the 76 remaining cases with notable changes~(17.0\%), authors added leakage checks, annotation workflows, and safety evaluations; introduced consent or review by institutional review boards; removed identifiers; or changed data access or release. For a process intended to steer research towards safer practice, substantive changes in only 17.0\% are concerning. Why do authors' revisions so often stop short of the implicated research itself?

\section{Qualitative Analysis: What Do Authors Change Instead?}
The previous section highlights what authors did or did not change, but not necessarily whether or not those choices were driven by reviewers' criticisms. To distinguish these possibilities, we read complete review histories for 25 cases and interview corresponding authors about their rebuttal and resubmission decisions. We select cases to cover different relationships between the flag and the resubmitted project: 11 apparent research changes, 6 pre-existing practices, and 8 cases with no apparent change, spanning harmful applications~(4), fairness and discrimination~(9), responsible research practice~(6), and privacy, security, and safety~(6). Of 25 researchers we invited, three responded; we refer to them as P1--P3. In each interview, we asked what they changed, what they paid most attention to, and whether reviewer feedback affected research decisions or only presentation.\footnote{After internal review, we ran 30-minute interviews under standard consent/privacy safeguards starting Apr. 15, 2026.} To avoid priming participants about why revisions occur, we keep the recruitment email neutral:\\
{\scriptsize
\textbf{Recruitment email~(abridged).} 
Hello \{First name\}, We are conducting a study on how ethics reviews at AI conferences shape papers after reviewers raise ethical concerns and a paper is rejected. We would appreciate a brief call to learn about your rebuttal and resubmission experience, and what, if anything, differs between the ICLR submission and the public version.
}
\vspace{-0.25em}

First, we notice that in the pool of 25 review histories, authors often respond at length to ethics criticism during rebuttal, including partial concessions absent from later resubmissions. Yet in most cases only one of three to five reviewers raises the concern; the others discuss technical issues. In at least one case, the reviewer emphasizing ethics scored the paper 6 while reviewers who did not flag ethics scored it 4--5. Several rebuttals frame the concern as a dispute over disciplinary conventions or required background rather than a valid reason to change the research.

Despite this resistance on substantive grounds, authors do revise their prose for resubmission. We call this pattern \emph{filtering}: adapting papers for publication without treating criticism as grounds to redirect research. Participant 1~(P1) describes resubmissions as ``mostly similar,'' with revisions focused on ``adding another table,'' ``more metrics,'' and issues that were ``mostly metric based.'' P1 further adds that peer reviews are ``mostly to filter out papers'' since many reviewers repeat objections to text the authors have already revised. Participant 2~(P2) says reviewers were ``very focused on a particular aspect, [...] instead of seeing a larger picture.'' The authors ``decided to just cut it out completely,'' rewrote the introduction for an audience in computer science, and recruited 5,000 participants for another annotation pass. Although P2 accepts several concerns as ``fair points,'' P2 calls the central criticism ``really about my presentation'' and peer review ``really about an editorial process.'' P2 attributes the narrowing to reviewers in computer science, saying that ``the CS reviewers chose to filter it'' and that reviewers ``have so much power shaping what stories get seen.''

Participant 3~(P3) reports rewriting the introduction and abstract while leaving methods and results essentially unchanged, and calls review closer to a ``coin flip'' than a reason to change the research. These publication changes can still be valuable: authors may narrow claims, clarify limitations, document provenance, or adopt safer release plans. But if conferences want ethics review to govern research, it should encourage longitudinal and deeply procedural changes that make research and its downstream consequences materially safer, not only improve what reaches publication. How can conferences make those changes more likely to persist?

\section{Discussion: How Should AI Research Be Governed?}
\textbf{Carry ethics flags across resubmissions.} When a paper is rejected, the original reviewers lose leverage over its subsequent trajectory. A new reviewer panel may never learn of the prior ethics concern and, given reviewer noise, may not re-raise it independently. This is exacerbated by our observation that ethics concerns are often concentrated in one reviewer rather than shared across the panel. Conferences should therefore require authors to confidentially disclose prior ethics flags on resubmission, including the category and whether they changed the research or disputed the concern. The new venue could ask an independent ethics reviewer to assess the concern without revealing the previous review, identities, scores, venue, or manuscript to reviewers or area chairs. This keeps unresolved concerns under independent scrutiny without binding the new venue to the prior verdict.

\textbf{Establish early-stage ethics review.} Feedback should arrive while implicated choices can still change. By full-paper review, data collection and experiments may already be complete; conferences could therefore require an early-stage ethics pre-submission as a prerequisite for later full-paper submission. \citet{stilgoe2013responsible} similarly argue for considering consequences before choices become difficult to revise. Such early review could scrutinize data collection and experimental design, while later review remains available for broader critiques.

\textbf{Improve reviewer fit.} Earlier feedback cannot help if authors regard criticism as a misunderstanding. Indeed, P1 discounts a reviewer who repeats an objection to removed text, while P2 says computer science reviewers focused on one aspect instead of ``seeing a larger picture.'' Better domain matching could make reviewers' criticism more credible~\citep{prunkl2021institutionalizing}.

\section{Limitations and Conclusion}

\textbf{Limitations.} Since we focus on papers we can confidently say are resubmissions, we may miss important cases wherein authors entirely abandoned their projects because of a flag, and therefore underestimate the impact of steering. Furthermore, our corpus includes only ICLR submissions, not NeurIPS or ICML. However, given the similar review--rebuttal--decision cycle across top AI conferences, we expect the mechanisms we identify to extend beyond ICLR, although their prevalence may differ. Additionally, only three of 25 invited researchers responded; this is reasonable for a qualitative study, but more research is needed to make claims about broader researcher sentiments.

{\widowpenalties 1 0
\textbf{Conclusion.} The results of our mixed-method study, realized through tracking the evolution of papers and interviewing real authors, strongly suggest that substantial revisions to a manuscript do not necessarily mean that ethics review redirected the underlying research. The resubmission types, review histories, and interviews explain why: concessions can matter immediately while reviewers still influence the decision; after rejection, the original reviewers lose that leverage, and reviewer noise may keep the concern from resurfacing. As such, conferences should review earlier, match by domain, and above all require the disclosure of prior flags upon resubmission so rejection does not reset accountability for an unresolved ethics concern when the project moves to another venue.\par}

\clearpage
\bibliographystyle{plainnat}
\bibliography{references}
\end{document}